\documentclass[runningheads,a4paper]{llncs}

\usepackage{amssymb}
\usepackage{graphicx}
\usepackage{url}
\usepackage{xcolor}
\newcommand{\keywords}[1]{\par\addvspace\baselineskip
\noindent\keywordname\enspace\ignorespaces#1}
\usepackage{multirow}
\usepackage{booktabs} 
\usepackage{algorithmic}
\usepackage{colortbl}
\usepackage[ruled]{algorithm2e} 

\usepackage{subfigure}
\usepackage{caption}

\newcommand{\diffblock}[1]{#1}

\urldef{\mailsa}\path|xiang.zhang3@student.unsw.edu.au,{lina.yao,chaoran.huang}@unsw.edu.au|
\urldef{\mailsb}\path|michael.sheng@mq.edu.au|
\urldef{\mailsc}\path|sandyawang@gmail.com|

\begin{document}

\mainmatter  

\title{Intent Recognition in Smart Living Through Deep Recurrent Neural Networks}

\titlerunning{Intent Recognition in Smart Living Through Deep Learning}

\author{Xiang Zhang$^1$ \and Lina Yao$^1$\and Chaoran Huang$^1$\and Quan Z. Sheng$^2$\and\\
Xianzhi Wang$^3$}
\authorrunning{Xiang Zhang\and Lina Yao\and Chaoran Huang et.al}
\institute{$^1$University of New South Wales, AU\\ $^2$Macquarie University, AU\\ $^3$Singapore Management University, Singapore\\
\mailsa\\
\mailsb\\
\mailsc\\
}

\maketitle

\begin{abstract}
Electroencephalography (EEG) signal based intent recognition has recently attracted much attention in both academia and industries, due to helping the elderly or motor-disabled people controlling smart devices to communicate with outer world.
 However, the utilization of EEG signals is challenged by low accuracy, arduous and time-consuming feature extraction. This paper proposes a 7-layer deep learning model
  to classify raw EEG signals with the aim of recognizing subjects' intents, to
avoid the time consumed in pre-processing and feature extraction. The hyper-parameters are selected by an Orthogonal Array experiment method for efficiency. Our model is applied to an open EEG dataset provided by PhysioNet and achieves the accuracy of 0.9553 on the intent recognition.
The applicability of our proposed model is further demonstrated by two use cases of smart living (assisted living with robotics and home automation).
\end{abstract}

\section{Introduction}
Smart living involves a collection of technologies that monitor and control domestic living environments, intended to support residents' routine activities to improve their quality of lives.
However, the existing smart living control technologies (e.g., voice control \cite{muhammad2017enhanced} and application-based control \cite{kumar2014ubiquitous}), may still be found difficult in situations that people have troubles in motor abilities, such as aged individuals, people having motor neuron disease(e.g., Parkinson disease, cord injury, brain-stem stroke) or disabilities.

Thus, to assist such individuals, new smart home systems based on intent recognition are essential, which likely can alleviate aforementioned issues.

Electroencephalography (EEG) signals reflect activities on certain brain areas not requiring any initiative actions such as gesture, voice, or so on. EEG data is generated when a subject imagines performing a certain action such as close hands. Therefore, EEG signal are widely captured to recognize one's intent, with the intent of using it as input to communicate or interact with external smart devices such as wheelchairs or service robots a real-time brain-computer interface (BCI) systems \cite{Alomari2014}.


So far, existing EEG-based intent recognition approaches face several challenges. First, the data pre-processing, parameters selection
 and feature engineering
 are time-consuming and highly dependent on human expertise. Second,
  current accuracies mostly center around 60 $\sim$ 85\% \cite{major2017effects,sun2016classification,shenoy2015shrinkage}, which are too low for real-world deployment. Finally, existing research mainly focus on
binary intents recognition while multi-intent scenario dominates the practical applications.

On the other hand, deep learning based approaches are capable of modelling high level representations as well as capturing complex relationships, which are often hidden in raw data, via stacking multiple layers of information processing modules in hierarchical architectures \cite{LeCun}.
Recurrent Neural Networks (RNNs) is one example
making use of sequential information. In particular, Long Short-Term Memory (LSTM) is one RNN architecture designed to model temporal sequences and their long-range dependencies, and often results in higher accurate compared to conventional RNNs \cite{sak2014long}. In this paper,
we propose a deep recurrent neural network model for intent recognition in smart living,
to help individuals with motor impairments.
Reusable source code and dataset are provided to reproduce the results\footnote{https://github.com/xiangzhang1015/EEG-based-Control}.
Our main contributions of this paper are highlighted as below:
\diffblock{
\begin{itemize}
\item We propose a LSTM recurrent neural network for smart living intent recognition, which directly processes raw EEG data under multi-class scenario.
\item We apply Orthogonal Array experiment method for hyper-parameters tuning, which saves 98.4\% of time compared to exhausting tuning.
\item We evaluate our approach over an open EEG dataset and achieves 0.9553 of accuracy. We also demonstrate the applicability of proposed intent recognition in two real use cases.
\end{itemize}
}

\section{Related Work}
The current application of EEG signals is mainly in medicine and neurology.
 \cite{page2015flexible} proposes a Logistic Regression (LR) approach to analyse EEG signals to detect seizure patient and achieves as high as 91\% of accuracy.
Wavelet analysis \cite{albert2016automatic} is employed to carry on a diagnosis of Traumatic Brain Injury (TBI) by quantitative EEG (qEEG) data and reaches 87.85\% of accuracy.
Power spectral density \cite{al2017predicting} are extracted as EEG data features to input into SVM, extreme learning machine and linear discriminant analysis to predict the outcome of Transcranial direct current stimulation (TDCS) treatment. The work achieves 76\% accuracy with the data from FC4 $\sim$ AF8 channels and 92\% with the data from CPz $\sim$ CP2 channels.

All the aforementioned literature uses binary classification and extracts features in different areas manually.
Recent research focuses more on the performance comparison of different classifiers.
 \cite{An2014} builds one deep belief net (DBN) classifier for each channel and combines them through Ada-boost algorithm and classifies the left and right hand motor imagery.The work achieves average 83\% accuracy.
\cite{sun2016classification} adopts SVM as the classifier and achieves an average accuracy of 65\% with the input data being denoised by a wavelet denoising algorithm before power spectral density (PSD) feature selection. \cite{Ward2016} yields an accuracy of 80\% with the foundational universal background models (UBMs) classifier after the data is processed by I-vectors and Joint Factor Analysis (JFA). 
\cite{Tabar2017} combined convolutional neural networks (CNN) and stacked autoencoders (SAE) to classify EEG Motor Imagery signals and results 90\% accuracy. The application of related methods in smart living in relatively limited. As an example, \cite{mu2015design} uses high pass and low pass filter to reduce the noise signal interference and extracts EEG features by fisher distance. The switch control experiment results show that their approach achieves an accuracy of 86\%.

\section{The Proposed Approach}
\label{sec:methodology}
\vspace{-2mm}

\begin{figure}
\includegraphics[width=\linewidth]{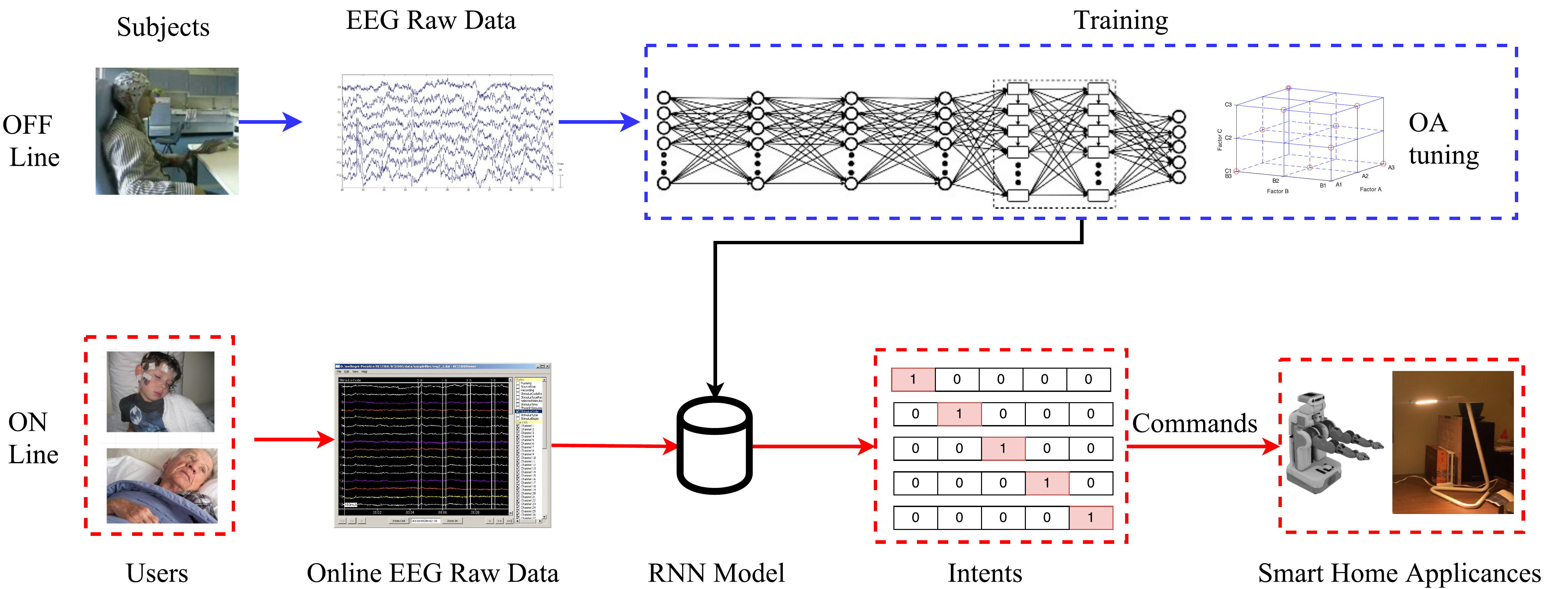}
\caption{Workflow of the Proposed Approach}
\vspace{-0.5cm}
\label{fig:overview}
\end{figure}
In this section we introduce the flow chart of the proposed approach at first and then involve to more details. 
The architecture of our approach is shown in Figure~\ref{fig:overview}. The system consists of two components: the online component and the offline component. 

In the online component, raw EEG data, collected from subjects,
are
used to train a deep recurrent neural network model (Section~\ref{sec:rnn}). The model directly works on raw EEG data without any pre-processing, smoothing, filtering or feature extraction. The parameters in the deep learning model are optimized by the Orthogonal Array experiment (Section~\ref{sec:oa}). In the offline component, the user's willing (EEG signal) is sent to above pre-trained RNN model and then recognized as specific intent. The intent is subsequently used to command devices, such as turning lights on/off or driving a robot to serve a cup of water.

\subsection{LSTM Recurrent Neural Network} 
\label{sec:rnn}
RNN, as a class of deep neural networks, can help to explore the feature dependencies over time through an internal state of the network, which allows us to exhibit dynamic temporal behavior.
In order to precisely recognize the user's intent in smart living surrounding, we propose a 7-layer LSTM Recurrent Neural Network model including three components: 1 input layer, 5 hidden layers, and 1 output layer. In hidden layers, two of them are consisted of LSTM cells \cite{zaremba2014recurrent} (shown as the rectangles in Figure~\ref{fig:overview}).

Assume one collection of EEG signals 
is 
$E = \{E_{1}, E_{2},..., E_{j}, ... , E_{bs}\}, E_{j} \in \mathbb{R}^K$ with $n_{bs}$ denotes the batch size, $j$ denotes the $j$-th EEG sample, and $K$ denotes the number of dimensions in each EEG raw signal ($K=64$ in this paper).
And in the RNN model, we denote the $i$-th layer ($i=1,2, \cdots, I$, $I=7$ in this paper) $X_i^r= \{X_{ijk}^r|k=1,2, \cdots, K_{i}\}, X_i^r \in \mathbb{R}^{[n_{bs},1,K_{i}]}$ ($K_1=K=64$), where $K_i$ denotes the dimension of the layer. Note that the number of dimension equals to the amount of neurons accordingly in each layer. When the input only contains one EEG sample, the first layer can be $X_1^r= E_{j}$.

Weights between layer \textit{i} and layer \textit{i+1} can be denoted as \(W_{i,(i+1)}^r \in \mathbb{R}^{[K_i,K_{i+1}]}\), for instance, $W_{1,2}^r$ describes the weight between layer 1 and layer 2. $b_i^r \in \mathbb{R}^{K_i}$ denotes the biases of \textit{i}-th layer.
The connection between the $i$-th and $(i+1)$-th layer will be $X_{i+1}^r=X_{i}^r*W_{i,i+1}^r+b_i^r$.

Please note the sizes of $X_{i}^r$, $W_{i,i+1}^r$ and $b_i^r$ must match. For example, in Figure~\ref{fig:overview}, the transformation between H1 layer and H2 layer, the sizes of $X_3^r$, $X_2^r$, $W_{[2,3]}$, and $b_2^r$ are correspondingly $[n_{bs},1,K_3]$, $[n_{bs},1,K_2]$, $[K_2,K_3]$, and $[n_{bs},1]$.
The 5-th and 6-th layers here are LSTM layers, and they can be connected by:
\[f_i=sigmoid(T(X_{(i-1)j}^r,X_{(i)(j-1)}^r))\]
\[f_f=sigmoid(T(X_{(i-1)j}^r,X_{(i)(j-1)}^r))\]
\[f_o=sigmoid(T(X_{(i-1)j}^r,X_{(i)(j-1)}^r))\]
\[f_m=tanh(T(X_{(i-1)j}^r,X_{(i)(j-1)}^r))\]
\[c_{ij}=f_f\odot c_{i(j-1)}+f_i\odot f_m\]
\[X_{ij}^r=f_o\odot tanh(c_{ij})\]
where $f_i,f_f,f_o$ and $f_m$ represent the input gate, forget gate, output gate and input modulation gate
accordingly, and $\odot$ denotes the element-wise multiplication. The $c_{ij}$ denotes the state (memory) in the $j$-th LSTM cell in the $i$-th layer, which is the most significant part to explore the time-series relevance between samples. The $T(X_{(i-1)j}^r,X_{(i)(j-1)}^r)$ denotes the operation as follows:
$$X_{(i-1)j}^r*W+X_{(i)(j-1)}^r*W'+b$$
where $W$, $W'$ and $b$ denote the corresponding weights and biases.
At last, we
obtain the RNN predict results $X_7^r$ and employ the cross-entropy as the cost function. The $\ell_2$ norm is selected as the regularization function and the cost is optimized by the AdamOptimizer algorithm \cite{kingma2014adam}.



\subsection{Orthogonal Array Experiment Method}
\label{sec:oa}
Although deep learning algorithms can generally achieve good performance in many areas, tuning the hyper-parameters (e.g., the number of layers, the number of nodes in each layer and the learning rate) is time-consuming and dependent on one's experience. This paper employs the Orthogonal Array (OA) experiment method \cite{taguchi1987system} to select the hyper-parameters, which works \emph{much faster} than traditional hyper-parameters tuning methods. OA\footnote{\url{https://www.york.ac.uk/depts/maths/tables/taguchi_table.htm}} is widely used in {\it design of experiments, coding theory, and cryptography}, however, to our best knowledge, this paper is the very first work to apply OA of the parameter tuning in machine learning and data mining areas.

\begin{figure}[ht]
\centering
\begin{minipage}[b]{0.4\linewidth}
\centering
\includegraphics[width=\textwidth]{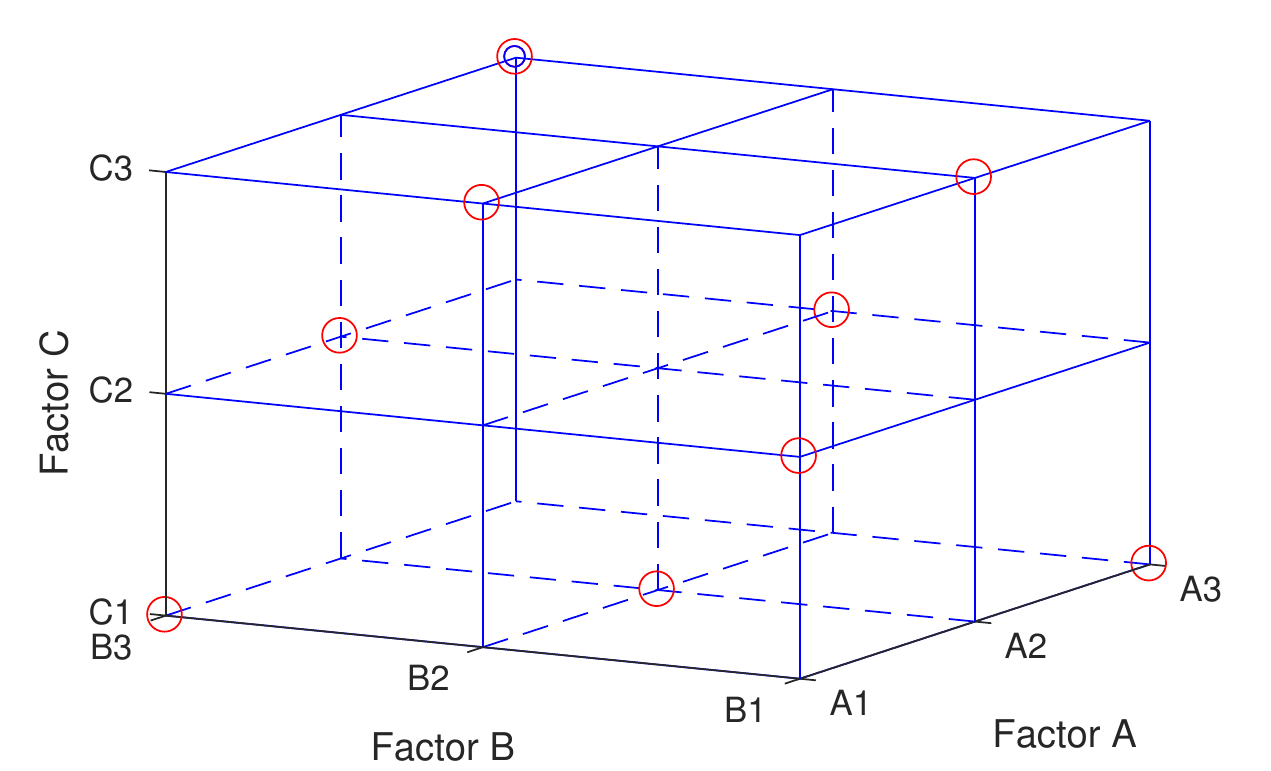}
\caption{OA selection}
\label{fig:figure5}
\end{minipage}
\hspace{2mm}
\begin{minipage}[b]{0.5\linewidth}
\centering
\resizebox{\linewidth}{!}{
\begin{tabular}{lcccc}
\hline
\rowcolor[HTML]{C0C0C0}
          & {\bf Level 1} & {\bf Level 2} & {\bf Level 3} & {\bf Level 4} \\ \hline
$\lambda$ & 0.002   & 0.004   & 0.006   & 0.008   \\
$lr$      & 0.005   & 0.01    & 0.015   & 0.02    \\
$K_i$     & 16      & 32      & 48      & 64      \\
$I$       & 5       & 6       & 7       & 8       \\
$n_b$      & 1       & 3       & 6       & 13      \\ \hline
\end{tabular}
}
\captionof{table}{Factors and levels}
\label{tab:factor}
\end{minipage}
\end{figure}

OA is a systematic and statistical method and its principle is to compare the dependent variable which is resulted from a different combination of independent variables. It chooses certain representative combinations instead of all combinations for testing. In this method, independent variable is called ``factor" and different values of factor are called ``levels".
For instance, if the program has three factors and each of them has three levels, which are represented by a cube with 27 nodes (each node represents one combination of hyper-parameters), OA only chooses 9 representative groups of parameters to optimize the selection. As shown in
Figure~\ref{fig:figure5}, $A_1, A_2, A_3$ represent 3 levels of factor $A$, while factors $B, C$ are by the same token (the factor is supposed to be statistically independent with the others). The 9 circled nodes are the nine groups selected by OA. Each edge (totally 27 edges) in the cube has one circled node and each face (totally 9 faces) has three circled nodes.

For different number of factors and levels, corresponding OA table is provided. Generally, an OA table can be written as $L{n_a} (n_b^{n_c})$, where $n_a$ denotes the number of hyper-parameter combination, $n_b$ denotes the number of levels of each factor and $n_c$ denotes the number of factors.

\section{Experiments}
\subsection{Dataset}
We select the widely used EEG data from PhysioNet eegmmidb (\textit{EEG
motor movement/imagery database}) database\footnote{\url{https://www.physionet.org/pn4/EEGmmidb/}}
  to evaluate the proposed approach.
The EEG signals we selected are under 5 categories of intents. The intents are shown in Table~\ref{tab:table2}.
In our work, we select 280,000 EEG samples from 10 subjects (28,000 samples each subject) for the experiment. Every sample is a vector of 64 elements corresponding to 64 channels.
\vspace{-3mm}
\begin{figure}
\centering
\includegraphics[width=0.8\textwidth]{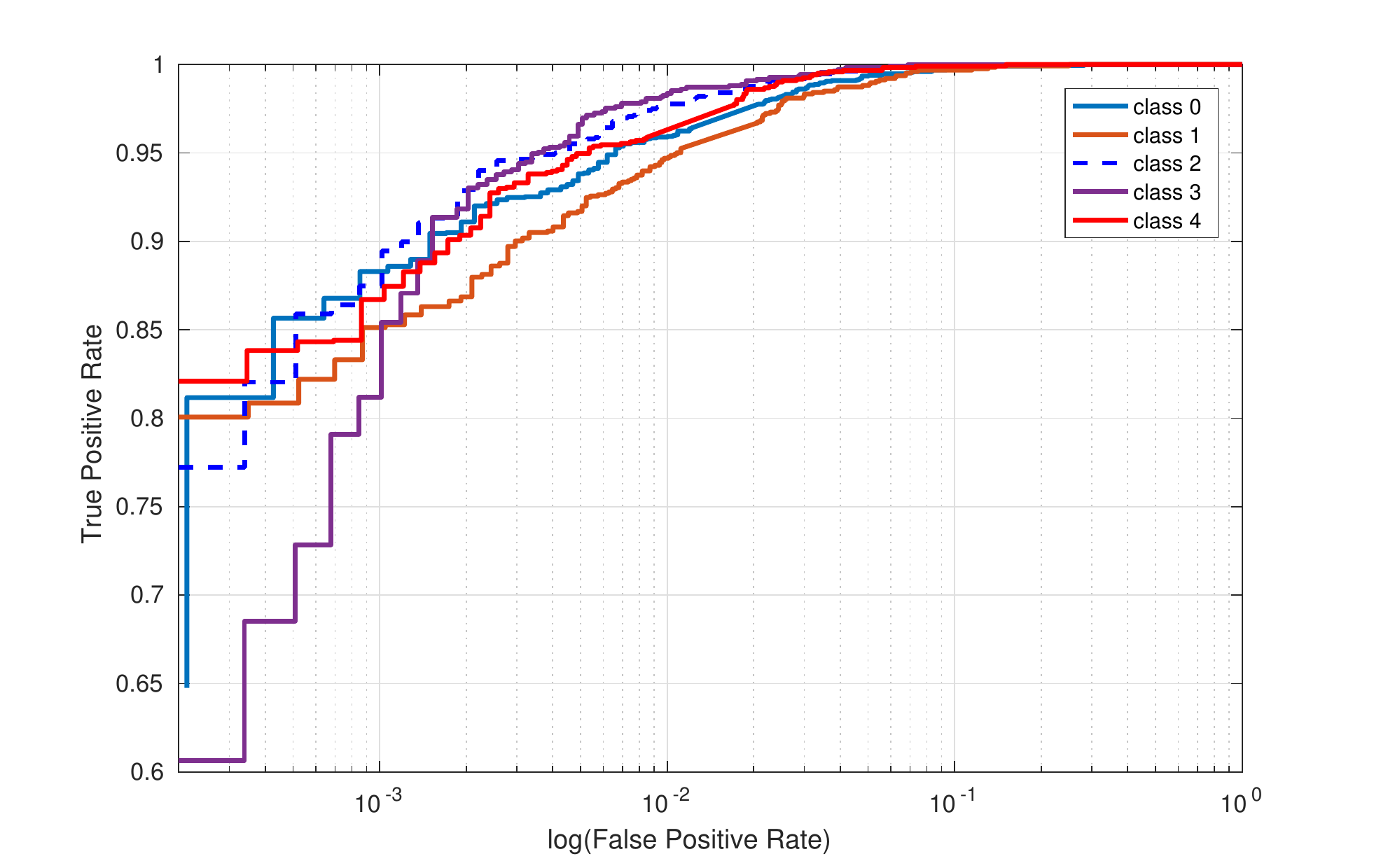}
\caption{ROC curves. X-axis is the logarithmic of the False Positive Rate.
 }
\label{fig:roc}
\end{figure}
\vspace{-3mm}

\vspace{-2mm}
\begin{table}[ht]
\begin{minipage}[b]{0.55\linewidth}
\centering
\resizebox{\linewidth}{!}{\begin{tabular}{lcccl}
\hline
\rowcolor[HTML]{C0C0C0}
{\bf Intent} & {\bf Label} & {\bf Robot (Case 1)} &{\bf Household Appliance (Case 2)} \\ \hline
Eye Closed         & 1                   & Walk Ahead           & Turn on Blue LEDs        \\
Left Fist    & 2                   & Turn Left                & Turn on White LED          \\
Right Fist   & 3                   & Turn Right               & Turn on Yellow LED       \\
Both Fists  & 4                   & Grasp                    & Turn on Red LED         \\
Both Feet    & 5                   & Unloose                  & Turn on All LEDs         \\ \hline
\end{tabular}
}
\caption{Intents and corresponding label and function in case studies}
\label{tab:table2}
\end{minipage}
\hspace{3mm}
\begin{minipage}[b]{0.44\linewidth}
\centering
\begin{scriptsize}
\resizebox{\linewidth}{!}{\begin{tabular}{lllllll|llll}
\hline
\rowcolor[HTML]{C0C0C0}
 &  & \multicolumn{5}{l|}{\cellcolor[HTML]{C0C0C0}\textbf{Ground Truth}} & \multicolumn{4}{l}{\cellcolor[HTML]{C0C0C0}\textbf{Evaluation}} \\ \hline
 &  & 0 & 1 & 2 & 3 & 4 & Precision & Recall & F1 & AUC \\
 & 0 & \cellcolor[HTML]{FE0000}2062 & 19 & 23 & 18 & 22 & 0.9618 & 0.9380 & 0.9497 & 0.9982 \\
 & 1 & 17 & \cellcolor[HTML]{34FF34}1120 & 19 & 15 & 20 & 0.9404 & 0.9084 & 0.9241 & 0.9977 \\
 & 2 & 13 & 13 & \cellcolor[HTML]{F8FF00}1146 & 14 & 11 & 0.9574 & 0.9257 & 0.9413 & 0.9990 \\
 & 3 & 10 & 5 & 7 & \cellcolor[HTML]{96FFFB}1162 & 10 & 0.9732 & 0.9028 & 0.9367 & 0.9990 \\
\multirow{-6}{*}{\begin{tabular}[c]{@{}l@{}}Predicted\\ Label\end{tabular}} & 4 & 18 & 21 & 15 & 23 &\cellcolor[HTML]{329A9D} 1197 & 0.9396 & 0.9392 & 0.9394 & 0.9987 \\
Total &  & 2120 & 1178 & 1210 & 1232 & 1260 & 4.7723 & 4.6140 & 4.6911 & 4.9926 \\
Average &  &  &  &  &  &  & \textbf{0.9545} & \textbf{0.9228} & \textbf{0.9382} & \textbf{0.9985} \\ \hline
\end{tabular}
}
\end{scriptsize}
\caption{The confusion matrix of 5-classes classification}
\label{tab:table5}
\end{minipage}
\vspace{-10mm}
\end{table}

\subsection{Overall Comparison}
\label{sub:comparison}
This section is aimed to demonstrate the efficiency of the proposed approach, for which we compare our approach with the state-of-the-art methods. Our model is composed of 7 layers RNN with 2 LSTM layers, the learning rate and the $\lambda$ are set as 0.004 and 0.005, the number of the nodes in each hidden layer is 64 and the number of batches $n_b$ is 3 (detailed in Section~\ref{sub:HT}).

Our intent recognition result, the confusion matrix and the corresponding evaluation are presented in Table~\ref{tab:table5}. It can be read that our approach produces a mean accuracy of \textbf{0.9553}, in tests of five intents recognition on 10 subjects. The ROC (Receiver Operating Characteristic) curves of five intents are displayed in Figure~\ref{fig:roc}.
Additionally, comparison with the state-of-the-art methods is shown in Table~\ref{tab:table6} (the Binary/Multi column refers binary intents recognition or multi-intents recognition). The KNN sets the number of neighbors as 3; the SVM adopts One-vs-the-rest (OvR) multi-class strategy and the estimator is LinearSVC; the RF sets the number of estimators as 300; the AdaBoost adopts the number of estimators as 50 and the learning rate as 0.3; all the not mentioned parameters are set as default values. We can perceive that the proposed approach significantly outperforms all the state-of-the-art methods, by a large margin of 10\%.

\subsection{Hyper-parameter Tuning}
\label{sub:HT}
The intent recognition results rely on hyper-parameters since we adopt deep learning model. To achieve optimal recognition accuracy, we employ OA to optimize the hyper-parameters. In this paper, we select five most common hyper-parameters including $\lambda$ (the coefficient of $\ell_2$ norm), $lr$ (learning rate), $K_i$(the hidden layer nodes size),  $I$ (the number of layers), and  $n_b$ (the number of batches\footnote{The size of training dataset and testing dataset depends on $n_b$ since the total dataset is fixed, e.g., if $n_b$ equals 1,  there will be 14,000 training dataset and 14,000 testing dataset. If $n_b$ equals 3, we will have 21,000 training dataset and 7,000 testing dataset}), and they are shown in Table~\ref{tab:factor}.
Since this OA experiment contains 5 factors and 4 levels, the total number of factor combinations can be found in \textit{the standard orthogonal experiment table}\footnote{\url{https://www.york.ac.uk/depts/maths/tables/l16b.htm}}. As shown in the standard orthogonal experiment table, 5 factors with 4 levels OA experiment has 16 different combine ways, which means 16 experiments should be conducted to optimize the hyper-parameters. The combination of hyper-parameters and the range analysis of results of the experiment, are shown in
Table~\ref{tab:table4}. The optical $\lambda$, $lr$, $K_i$, $I$, and $n_b$ tuned by OA are 0.004, 0.005, 64, 7, and 3, respectively. The parameter selection of 5 factors and 4 levels needs $1024=4^5$ combinations in an exhaustive method, while with OA only 16 combinations are needed. This means $(1-16/1024)=\textbf{98.4\%}$ 
of
time are saved.
In Table~\ref{tab:table4}, $R_{leveli}$ is the sum of accuracy of all the combinations contains $level_i$.
We selected the best levels listed in Table~\ref{tab:table4} for training the model and obtain an accuracy of \textbf{0.9553}.

\vspace{-5mm}
\begin{table}[htbp]
\centering
\caption{OA experiment factor analysis}
\label{tab:table4}
\resizebox{\linewidth}{!}{
\begin{tabular}{lllllllllllllllllllllcllllll}
\hline
\rowcolor[HTML]{C0C0C0}
\textbf{No.} & \textbf{1} & \textbf{2} & \textbf{3} & \textbf{4} & \textbf{5} & \textbf{6} & \textbf{7} & \textbf{8} & \textbf{9} & \textbf{10} & \textbf{11} & \textbf{12} & \textbf{13} & \textbf{14} & \textbf{15} & \textbf{16} & \textbf{$R_{level1}$} & \textbf{$R_{level2}$} & \textbf{$R_{level3}$} & \textbf{$R_{level4}$} & \textbf{Best level} \\ \hline
$\lambda$ & 0.002 & 0.002 & 0.002 & 0.002 & 0.004 & 0.004 & 0.004 & 0.004 & 0.006 & 0.006 & 0.006 & 0.006 & 0.008 & 0.008 & 0.008 & 0.008 & 3.159 & \textbf{3.26} & 2.441 & 2.44 & \textbf{0.004} \\
$lr$ & 0.005 & 0.01 & 0.015 & 0.02 & 0.005 & 0.01 & 0.015 & 0.02 & 0.005 & 0.01 & 0.015 & 0.02 & 0.005 & 0.01 & 0.015 & 0.02 & \textbf{3.47} & 2.875 & 2.747 & 2.208 & \textbf{0.005} \\
$K_i$ & 16 & 32 & 48 & 64 & 32 & 16 & 64 & 48 & 48 & 64 & 16 & 32 & 64 & 48 & 32 & 16 & 2.132 & 2.886 & 3.011 & \textbf{3.271} &\textbf{ 64} \\
$I$ & 5 & 6 & 7 & 8 & 7 & 8 & 5 & 6 & 8 & 7 & 6 & 5 & 6 & 5 & 8 & 7 & 2.326 & 2.932 & \textbf{3.048} & 2.894 & \textbf{7} \\
$n_b$ & 1 & 3 & 6 & 13 & 13 & 6 & 3 & 1 & 3 & 1 & 13 & 6 & 6 & 13 & 1 & 3 & 2.969 & \textbf{3.088} & 2.907 & 2.336 & \textbf{3} \\
acc & 0.689 & 0.91 & 0.893 & 0.667 & 0.925 & 0.717 & 0.848 & 0.77 & 0.926 & 0.826 & 0.322 & 0.367 & 0.93 & 0.422 & 0.684 & 0.404 &  &  &  &  & \\ \hline
\end{tabular}
}
\vspace{-3mm}
\end{table}

\subsection{Feature Evolution}
\label{sub:evolution}
\begin{figure}
\centering
\subfigure[input layer]{
  \label{fig:rnn_lr}
  \includegraphics[width=0.22\linewidth]{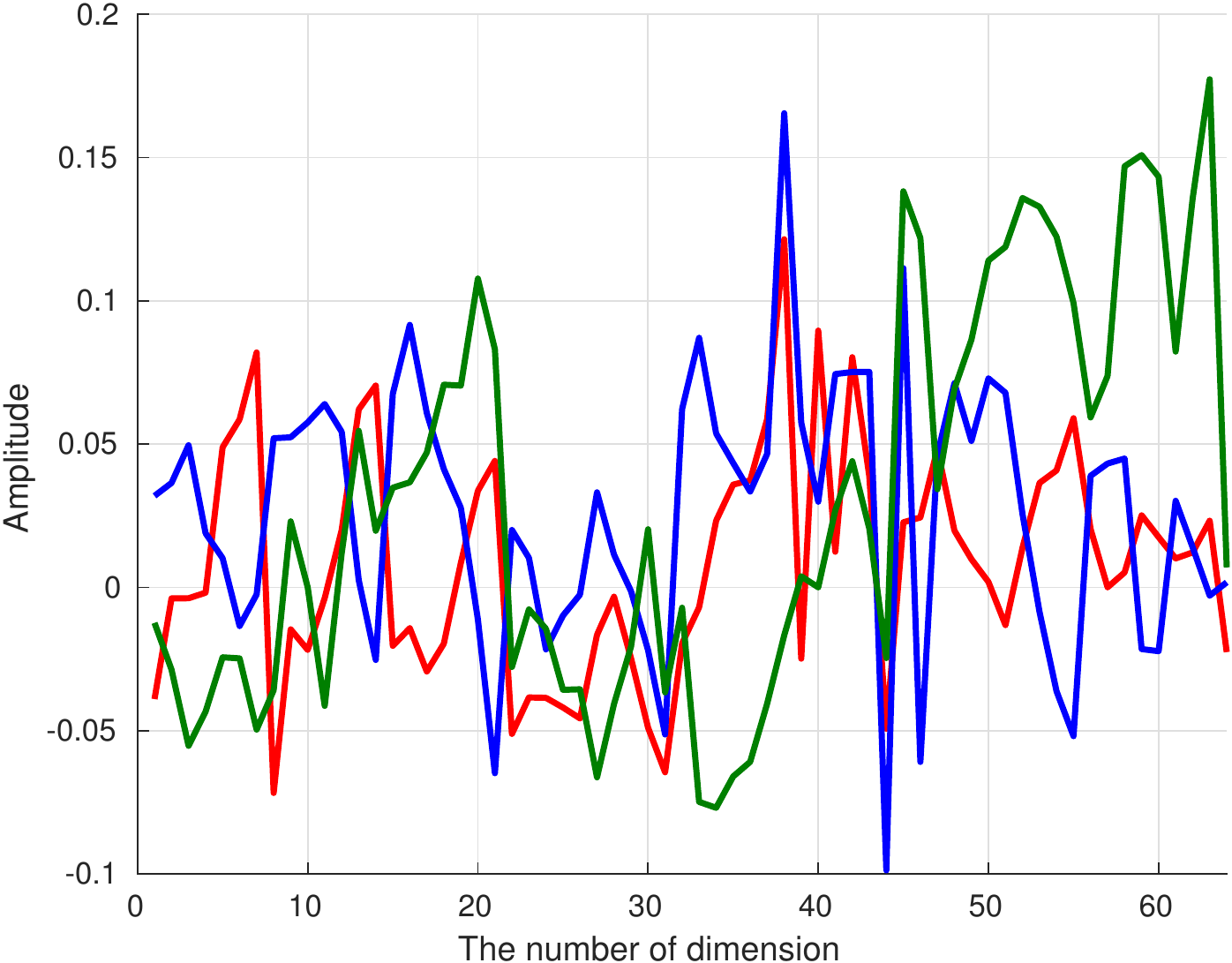}
}
\subfigure[hidden layer 1]{
  \label{fig:cnn_lr}
  \includegraphics[width=0.22\linewidth]{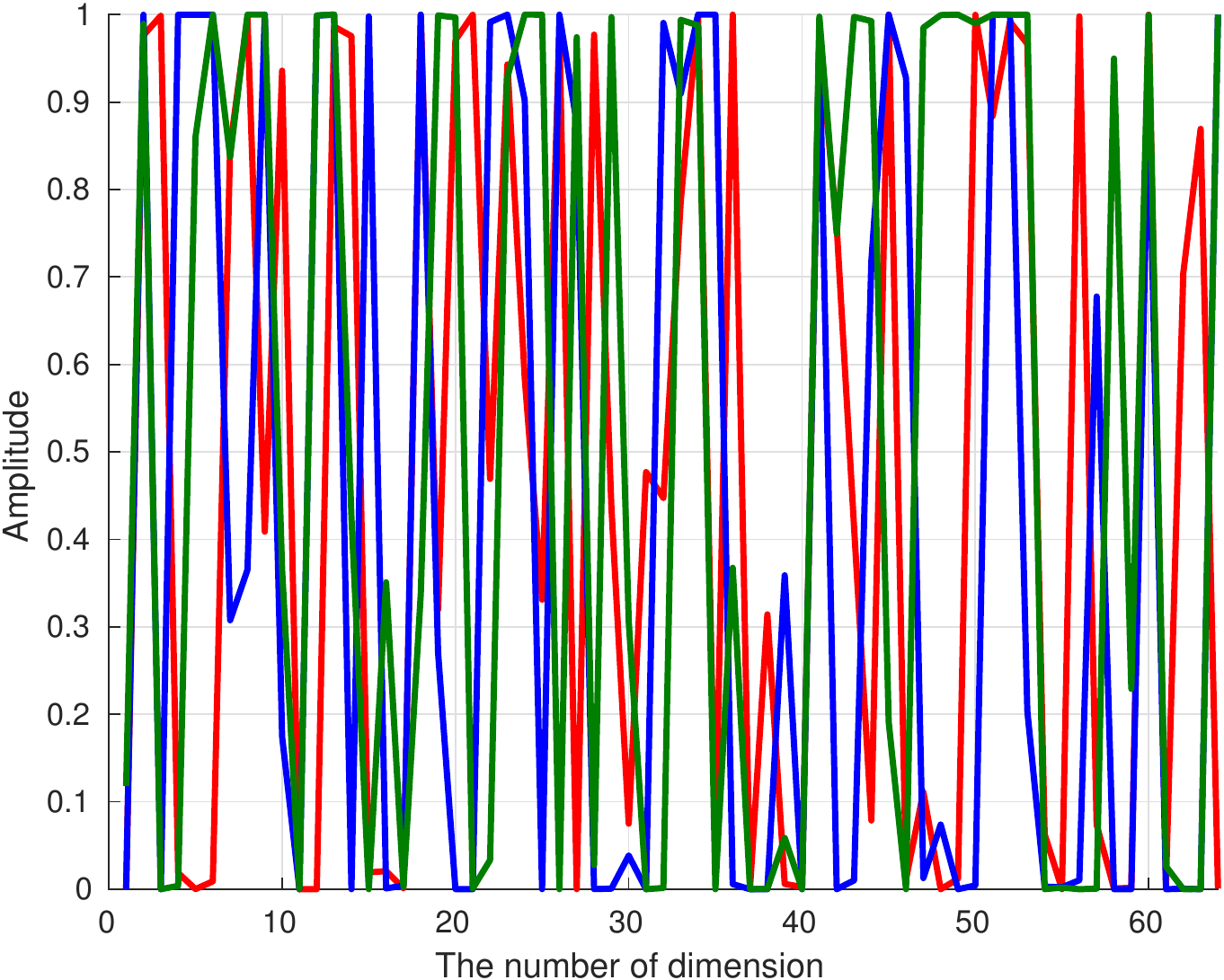}
}
\subfigure[hidden layer 3]{
  \label{fig:ae_lr}
  \includegraphics[width=0.22\linewidth]{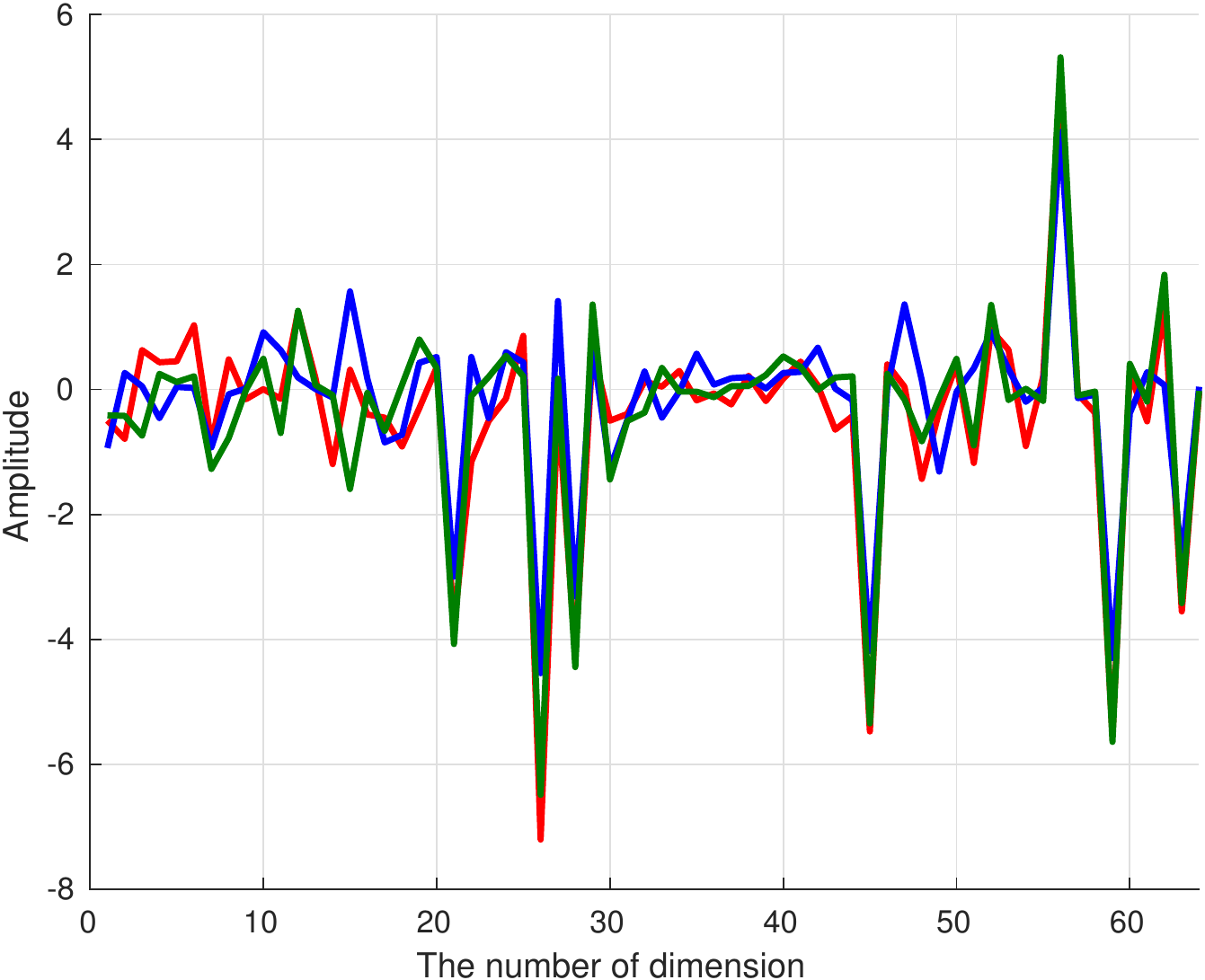}
}
\subfigure[LSTM layer 2]{
  \label{fig:xgb_lr}
  \includegraphics[width=0.22\linewidth]{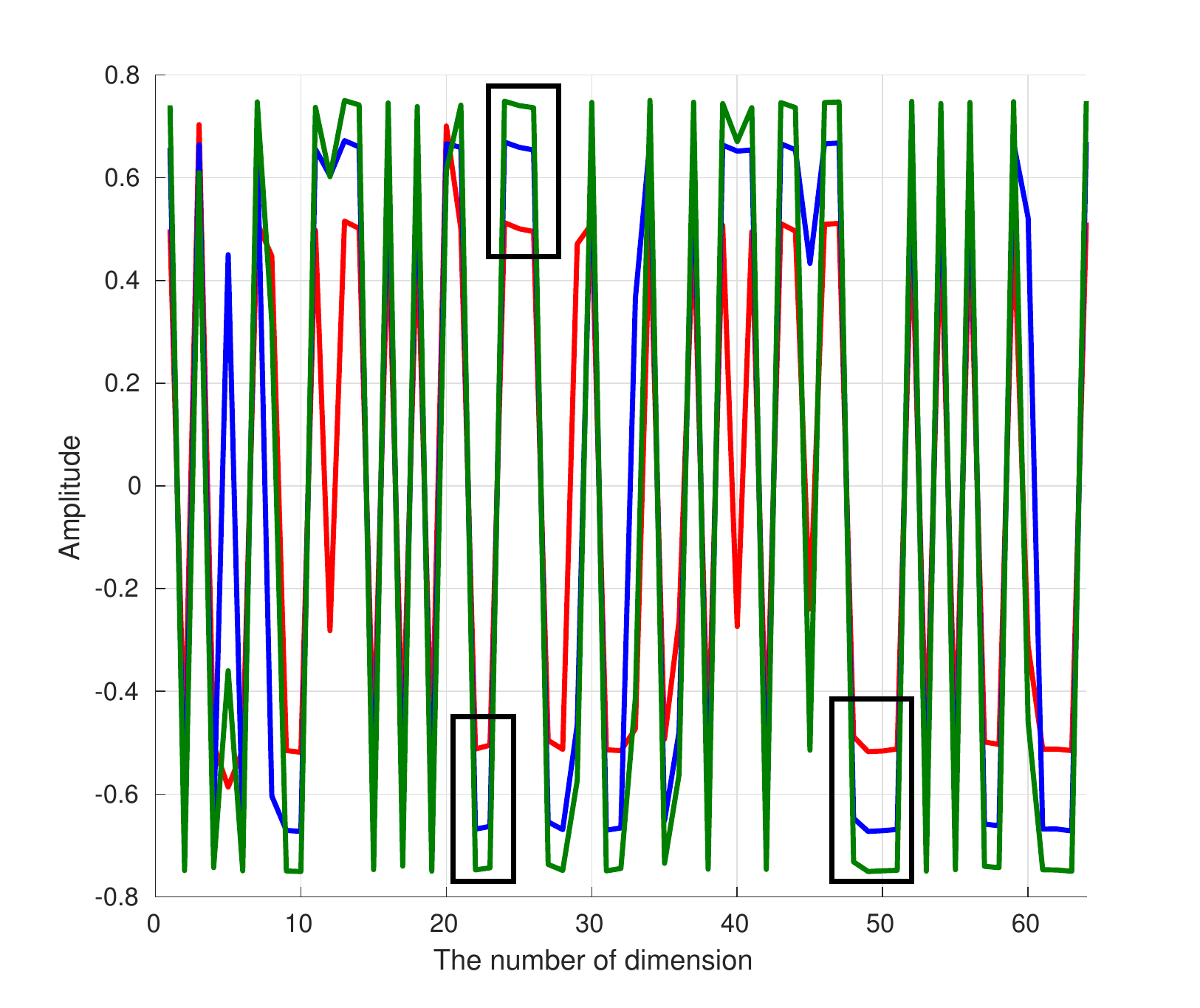}
}
\caption{Feature evolution. The black rectangles in Figure~\ref{fig:xgb_lr} indicate the features which can clearly show the difference between the various intents.}
\label{fig:feature_evolution}
\vspace{-10mm}
\end{figure}

To better understand the essence of the proposed model, we graphically describe the feature evolution procedures.
Figure~\ref{fig:feature_evolution} shows the revolution of variations between samples from different classes. In the input layer, the samples are chaotic entangled; and they become clear and observable in the last LSTM layer after the training through several hidden layers. Particularly, in Figure~\ref{fig:xgb_lr}, the black rectangles display parts of the dimensions which can clearly show the difference between the intents. Conclusively, the proposed approach is enabled to automatically extract distinguishable features (Figure~\ref{fig:xgb_lr}) from the chaotic raw EEG data (Figure~\ref{fig:rnn_lr}).
\vspace{-2mm}

\subsection{Deployment}
In this section, the efficiency of intent recognition is demonstrated by two applications. The structure of RNN and the corresponding parameters used in this section are the same as the counterparts in Section~\ref{sub:comparison}.
\subsubsection{ Assisted Living with Mind-controlled Mobile Robot}

A simulated robot is navigated by our system, which learns user's intent from EEG recordings, to take a can of beverage from a table in the kitchen and put it in a table in living room. 
  This case randomly selects some EEG raw data from Subject 1 dataset as simulation inputs. The path is shown in Figure~\ref{fig:figure7}, which is designed for the EEG data to drive PR2 to implement its service task. Starting from near the Kitchen's table, the PR2 robot walks forward and holds its hand to grasp the beverage can. Then it turns back and walks along the path to the table in living room and unlooses hands to put the beverage on the table.
 It shows that the robot can precisely grasp and unloose target according to the path planned in the subject's mind.
  The simulation platform is in Gazebo toolbox\footnote{\url{http://gazebosim.org/}} and the robot controlling program is powered by Robot Operating System (ROS)\footnote{\url{http://www.ros.org/}}.
  The simulation environment is depicted in Figure~\ref{fig:figure7} and the demo can be found at here\footnote{\url{https://www.youtube.com/watch?v=VZYX1095Vkc}}. The robot executes 5 actions according to 5 commands described in Table~\ref{tab:table2}.

\subsubsection{Assisted Living with Mind-controlled Appliances}
The most common scenario in a smart home would be controlling household appliances. In this case, we control four LEDs ON/OFF through intents.
LED commands corresponding to specific intents are mentioned in Table~\ref{tab:table2}.
For every command, the corresponding LED keeps on for 2 seconds and then turns off.
Such test is conducted 10 times with totally 80 commands, and our model accomplishes {\em 100\%} of accuracy, which indicates that the EEG-based mind control have potential to be significant in household in the future.

\vspace{-3mm}
\begin{table}[htbp]
\begin{minipage}[b]{0.55\linewidth}
\centering
\begin{scriptsize}
\resizebox{\linewidth}{!}{\begin{tabular}{lclcl}
\hline
\rowcolor[HTML]{C0C0C0}
 & \textbf{Index} & \textbf{Methods} & \textbf{Binary/Multi} & \textbf{Accuracy} \\ \hline
 & 1 & Almoari \cite{Alomari2014} & \multirow{5}{*}{Binary} & 0.7497 \\
 & 2 & Sun \cite{sun2016classification} &  & 0.65 \\
 & 3 & Major \cite{major2017effects} &  & 0.68 \\
 \begin{tabular}[c]{@{}l@{}}State\\ of the art\end{tabular}& 4 & Shenoy \cite{shenoy2015shrinkage} &  & 0.8206 \\
 & 5 & Tolic \cite{tolic2013classification} &  & 0.6821 \\
 & 6 & Ward \cite{Ward2016} & Multi (3) & 0.8 \\
 & 7 & Pinheiro \cite{pinheiro2016wheelchair} & Multi (4) & 0.8505 \\ \hline
\multirow{7}{*}{Baselines} & 8 & KNN (k=3) & \multirow{8}{*}{Multi (5)} & 0.8369 \\
 & 9 & SVM &  & 0.5082 \\
 & 10 & RF &  & 0.7739 \\
 & 11 & LDA &  & 0.5127 \\
 & 12 & AdaBoost &  & 0.3431 \\
 & 13 & CNN &  & 0.8409 \\
 & 14 & Ours &  & \textbf{0.9553} \\ \hline
\end{tabular}
}
\end{scriptsize}
\caption{Performance comparison with the state of the art methods. RF: Random Forest, LDA: Linear Discriminant Analysis. \textbf{All the methods are evaluated using the same database.}}
\label{tab:table6}
\end{minipage}
\hspace{2mm}
\begin{minipage}[b]{0.45\linewidth}
\includegraphics[width=\textwidth]{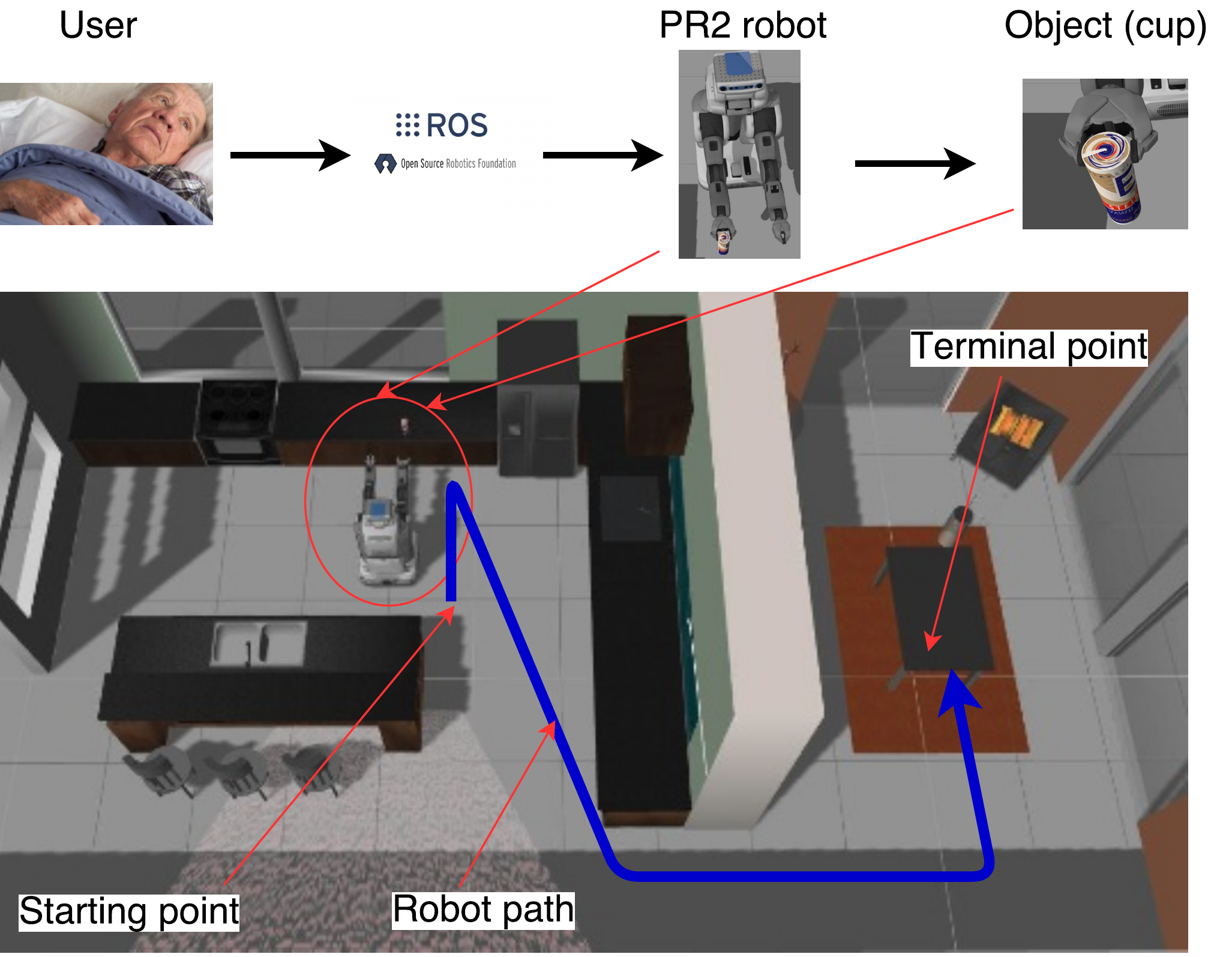}
\captionof{figure}{Use Case 1: mind-controlled PR2 assistive robot performs a daily task: reaching a cup of water in kitchen area and getting it back onto a table in living room.}
\label{fig:figure7}
\end{minipage}
\vspace{-15mm}
\end{table}

\section{Conclusion and Futurework}
In this paper, we present an LSTM-RNN approach to recognize the smart living user intents in EEG raw signals.
By experimenting on large scale EEG dataset,
we can claim that our proposed approach significantly outperforms a series of the state-of-the-art methods by achieving 0.9553 of accuracy.
It provides insight into feature revolution by visualizing the data shape, waveform fluctuation flowing through each layer of our proposed model.
Moreover, we demonstrate the applicability of the approach by implementing two use cases, wherein an assistive robot performs a physical task, and
household appliances are interacted, based on intent recognition.
Our prior work atop multi-task learning based framework \cite{yao2016learning} shows the capability to capture certain underlying local commonalities under the intra-class variabilities shared by all the activities of different subjects. Our future works will focus on improving the accuracy in {\em person-independent} scenario, wherein the training and testing data can be from different subjects.

\end{document}